\documentclass{PoS}

\usepackage{amsmath}
\usepackage{subfigure}
\usepackage{epsfig}

\newcommand{\rmR}{{\rm R}}
\newcommand{\rmI}{{\rm I}}
\newcommand{\rmMax}{{\rm max}}

\author{Gert Aarts and \speaker{Frank A.\ James}\\
        Department of Physics,\ Swansea University, Swansea, UK\\ 
        E-mail: \email{g.aarts@swansea.ac.uk},
		\email{pyfj@swansea.ac.uk}}
\title{The XY model at finite chemical potential using complex Langevin dynamics}
\ShortTitle{XY model at finite chemical potential}
\abstract{The three-dimensional XY model is studied at finite chemical potential using complex Langevin dynamics.
An adaptive stepsize algorithm is implemented to cure the problem of runaway solutions that appears when using a constant
stepsize. The validity of complex Langevin dynamics is tested against calculations using imaginary chemical potential and the world line formalism. 
While complex Langevin dynamics is found to work correctly at larger $\beta$, it fails for smaller $\beta$, in the region
of the phase diagram corresponding to the disordered phase. Diagnostic tests are developed to identify behaviour symptomatic of incorrect
convergence. These indicate that the failure does not depend on the severeness of the sign problem, but has a different origin.} 
\FullConference{The XXVIII International Symposium on Lattice Field Theory, Lattice2010\\
		June 14-19, 2010\\
		Villasimius, Italy}

\begin{document}
\section{Introduction}
Field theories with a complex action are difficult to treat nonperturbatively because the weight $e^{-S} = |e^{-S}|e^{i\varphi}$ in the 
partition function is not real. Standard numerical approaches based on a probability interpretation and importance sampling will
then typically break down, which is commonly referred to as the sign problem. This is a particularly pressing issue
with regards to the determination of the phase diagram of QCD in the plane of temperature and chemical potential. A 
comprehensive review can be found in Ref.~\cite{deForcrand:2010ys}

Complex Langevin dynamics offers the possibility of a general solution to this problem \cite{Parisi:1984cs, Klauder:1983, Damgaard:1987rr}. 
In this formulation, fields $\phi$ are 
supplemented with a fictional time-like dimension, $\vartheta$, and the system evolves according to the stochastic equation
\begin{equation}
\frac{\partial \phi_{x}(\vartheta)}{\partial \vartheta} = -\frac{\delta S[\phi;\vartheta]}{\delta \phi_{x}(\vartheta)} + \eta_{x}(\vartheta).
\end{equation}
In the case of a complex action, the fields are \emph{complexified} as $\phi \to \phi^\rmR + i\phi^\rmI$,
and the Langevin equations read (using real noise)
\begin{subequations}
\begin{align}
\frac{\partial \phi_{x}^\rmR}{\partial \vartheta} &= K^\rmR_{x} + \eta_{x},&\quad&
K_{x}^\rmR = -\mbox{Re}\left.\frac{\delta S}{\delta \phi_{x}}\right|_{\phi\to\phi^\rmR+i\phi^\rmI},\\
\frac{\partial \phi_{x}^\rmI}{\partial \vartheta} &= K^\rmI_{x},&\quad&
K_{x}^\rmI = -\mbox{Im}\left.\frac{\delta S}{\delta \phi_{x}}\right|_{\phi\to\phi^\rmR+i\phi^\rmI} ,
\end{align}
\end{subequations}
where the noise is Gaussian, 
$\langle \eta_{x}(\vartheta)\rangle = 0$, $\langle \eta_{x}(\vartheta)\eta_{y}(\vartheta^\prime)\rangle = 2\delta_{xy}\delta(\vartheta - \vartheta^\prime)$.
In the limit that $\vartheta\to\infty$ noise averages should become equal to quantum expectation values. 
Since the action is used to compute the drift terms, but not for any importance sampling, complex Langevin dynamics
can potentially avoid the sign problem. 

\section{XY model}
Motivated by previous studies of complex Langevin dynamics for QCD in the heavy dense limit \cite{Aarts:2008rr}, 
we consider the XY model at finite chemical potential \cite{Aarts:2010aq}.
This theory is closely related to the Bose gas, for which complex Langevin dynamics is known to work well \cite{Aarts:2008wh, Aarts:2009hn}. 
The XY model at finite chemical potential has the action 
\begin{equation}
S = -\beta\sum_{x}\sum_{\nu=0}^{2}\cos(\phi_{x} - \phi_{x+\hat{\nu}} - i\mu\delta_{\nu,0}) ,
\end{equation}
where $0\leq\phi_{x}<2\pi$. The theory is defined on a lattice of volume $\Omega = N_{\tau}N_{s}^2$, with periodic boundary
conditions. The chemical potential is coupled to the Noether charge associated with the global symmetry, $\phi_{x}\to\phi_{x}+\alpha$, and
is introduced in the standard way \cite{Hasenfratz:1983ba}. The action satisfies $S^{*}(\mu) = S(-\mu^{*})$ and at vanishing chemical 
potential the theory is known to undergo a phase transition at $\beta_{c} = 0.45421$ \cite{Banerjee:2010kc} between a disordered phase when $\beta<\beta_{c}$
and an ordered phase when $\beta>\beta_{c}$. 
The drift terms in the complex Langevin equations are given by
\begin{subequations}
\begin{align}
K_{x}^\rmR = -\beta&\sum_{\nu}\left[\sin(\phi_{x}^\rmR- \phi_{x+\hat{\nu}}^\rmR)\cosh(\phi_{x}^\rmI - \phi_{x+\hat{\nu}}^\rmI - \mu\delta_{\nu,0})\right.\\
&\left.+\sin(\phi_{x}^\rmR- \phi_{x-\hat{\nu}}^\rmR)\cosh(\phi_{x}^\rmI - \phi_{x-\hat{\nu}}^\rmI + \mu\delta_{\nu,0})\right] ,\nonumber\\
K_{x}^\rmI = -\beta&\sum_{\nu}\left[\cos(\phi_{x}^\rmR- \phi_{x+\hat{\nu}}^\rmR)\sinh(\phi_{x}^\rmI - \phi_{x+\hat{\nu}}^\rmI - \mu\delta_{\nu,0})\right.\\
&\left.+\cos(\phi_{x}^\rmR- \phi_{x-\hat{\nu}}^\rmR)\sinh(\phi_{x}^\rmI - \phi_{x-\hat{\nu}}^\rmI + \mu\delta_{\nu,0})\right] .\nonumber
\end{align}
\label{eq:drift-terms}
\end{subequations}
By choosing an imaginary chemical potential $\mu=i\mu_{I}$, the action becomes purely real and standard algorithms can be applied (here real Langevin 
dynamics is used). The behaviour at $\mu^2\gtrsim0$ can then be assessed by continuation of the behaviour found at $\mu^2\lesssim0$. 

\section{Adaptive stepsize}
In order to integrate the equations, the Langevin time needs to be discretized as $\vartheta=\epsilon n$. When real Langevin dynamics
can be employed (e.g. when $\mu^2\leq0$), the drift terms are bounded and a fixed stepsize is sufficient. However, in the case of complex Langevin dynamics,
the drift terms are unbounded (see Eq.~\ref{eq:drift-terms}) and numerical instabilities are encountered when the forces become large. These cause the system 
to diverge along ``runaway'' solutions. 

This can be cured by adjusting the stepsize when the configuration approaches a divergent trajectory \cite{Aarts:2009dg}. 
The adaptive stepsize is implemented by monitoring the maximal force term,
\[ K^\rmMax_{n} = \max_{x} \left|K_{x}^\rmR(n) + i K_{x}^\rmI(n)\right| ,\]
and at each update defining the stepsize to be 
\begin{equation}
\epsilon_{n} = \min\left\{\bar{\epsilon}, \bar{\epsilon}\frac{\langle K^\rmMax\rangle}{K^\rmMax_{n}}\right\} .
\end{equation}
Here $\bar{\epsilon}$ is the desired target stepsize and $\langle K^\rmMax\rangle$ is either precomputed or computed during the thermalisation
period. 

All observables are analyzed over equal periods of Langevin time and weighted with the stepsize to ensure correct statistical significance,
\begin{equation}
\langle O\rangle = \frac{\sum_{n}\epsilon_{n} O_{n}}{\sum_{n}\epsilon_{n}},
\end{equation}
with the total number of updates such that $\sum_{n}\epsilon_{n}$ is constant. 

While with a fixed stepsize it is practically impossible to generate a thermalised configuration, we found that
with an adaptive stepsize instabilities are completely eliminated. The same result applies to heavy dense QCD \cite{Aarts:2009dg}.

\section{World line formulation}
A useful feature of the XY model is that it can be rewritten exactly without a sign problem in terms of world lines. This dual formulation
can be efficiently simulated using a worm algorithm \cite{Banerjee:2010kc}. 
The partition function can be expanded into a sum over Bessel functions using the identity
\begin{equation}
e^{\beta\cos\phi} = \sum_{k=-\infty}^{\infty}I_{k}(\beta)e^{ik\phi},
\end{equation}
which allows the partition function to be written as
\begin{equation}
Z = \int D\phi\, e^{-S} = \sum_{[k]}\prod_{x,\nu}I_{k_{x,\nu}}(\beta)e^{k_{x,\nu}\mu\delta_{\nu,0}}\delta\left(\sum_{\nu}[k_{x,\nu} - k_{x-\hat{\nu},\nu}]\right).
\end{equation}
The sum over $[k]$ indicates a sum over all possible world line configurations. The action, 
$\langle S\rangle = -\beta\frac{\partial \ln Z}{\partial \beta}$, can be computed from
\begin{equation}
\langle S\rangle = -\beta\left\langle\sum_{x,\nu}\left[\frac{I_{k_{x,\nu}-1}(\beta)}{I_{k_{x,\nu}}(\beta)} - 
\frac{k_{x,\nu}}{\beta}\right]\right\rangle_{\mbox{\scriptsize{wl}}},
\end{equation}
where the average is taken over world line configurations. 

\section{Comparison}
\begin{figure}[htpb]
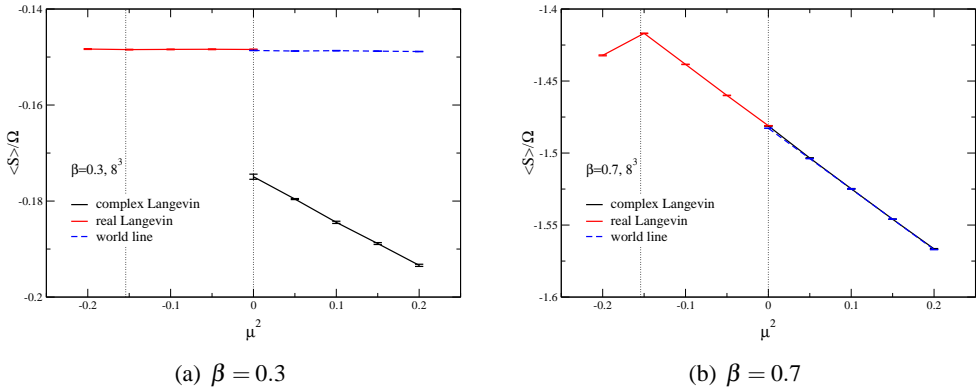

\centering
\subfigure[$\beta=0.3$]{
\includegraphics[totalheight=0.2\textheight, width=0.4\textwidth]{act_b0.3_x8.eps}}
\hspace{0.2in}
\subfigure[$\beta=0.7$]{
\includegraphics[totalheight=0.2\textheight, width=0.4\textwidth]{act_b0.7_x8.eps}}
\caption{Real part of the action density around $\mu^2\sim0$ at low and high $\beta$, with the world line result for comparison. 
The lattice volume is $8^3$. Note that with an imaginary chemical potential the theory is periodic under $\phi\to\phi+2\pi/N_\tau$ 
which yields a Roberge-Weiss transition at $\mu_{{\rm I}} = \pi/N_\tau$, similar to what is found in QCD.}
\label{fig:action-plots}
\end{figure}

To assess the validity of the results from complex Langevin simulations, a good starting point is to compare 
with results from imaginary chemical potential simulations at $\mu^2<0$. In Fig.~\ref{fig:action-plots} 
it can be seen that at large coupling $\beta=0.7$ the action density is continuous over the boundary at $\mu^2=0$ between
real and complex Langevin dynamics. The results at $\beta=0.3$ are in contrast to this. Here, the action density between the
two regions are not in agreement, exemplified by the fact that at $\mu^2=0$ the two simulations give different results 
for $\langle S\rangle /\Omega$. 
These observations are corroborated by the results from the world line formalism, which are consistent with the predictions 
from imaginary chemical potential calculations. 

\begin{figure}[htpb]
\centering
\vspace{-0.5in}
\includegraphics[totalheight=0.4\textheight, width=0.7\textwidth]{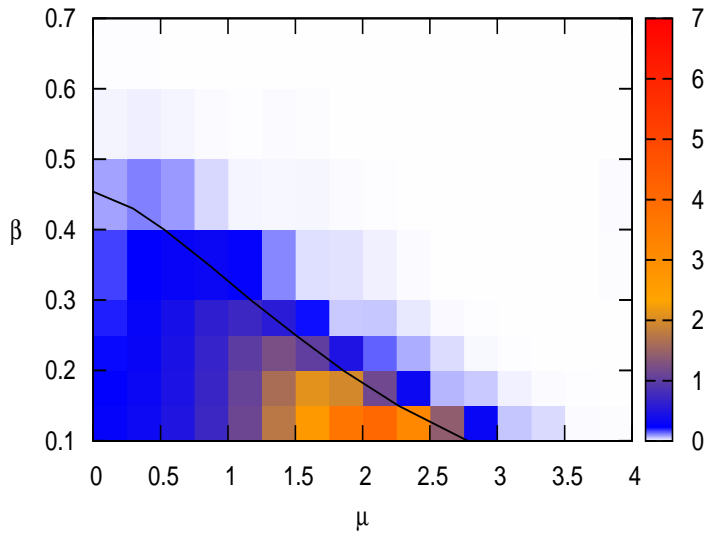}
\vspace{-0.4in}
\caption{Colour plot of relative difference $\Delta S$ between action density as computed using
complex Langevin dynamics and world line formulation. Also shown is the phase boundary $\beta_{c}(\mu)$
between ordered phase (large $\beta$) and disordered phase (small $\beta$).}
\label{fig:colour-plot}
\end{figure}

In order to investigate this further, we have studied a large number of parameter values in the $\beta$-$\mu$ plane. 
The disagreement between complex Langevin dynamics (cl) and world line formalism (wl) can be quantified by the relative
difference between the expectation values of the action, according to
\begin{equation}
\Delta S = \frac{\langle S\rangle_{{\rm wl}} - \langle S\rangle_{{\rm cl}}}{\langle S\rangle_{{\rm wl}}} .
\end{equation}
This is plotted in Fig.~\ref{fig:colour-plot} in the $\beta$-$\mu$ plane, with the phase boundary taken
from Ref.~\cite{Banerjee:2010kc}. It can be seen that the breakdown of complex Langevin dynamics is highly correlated with the phase
boundary of the theory: complex Langevin dynamics is in agreement in the ordered phase but begins to break down at the boundary
and is in complete disagreement in the disordered phase. 

\section{Diagnostics}
To identify the manner in which complex Langevin dynamics makes the transition from working correctly (at large $\beta$) to 
failing (at small $\beta$), a good test is to compare distributions of observables computed at $\mu=0$ using different 
initial conditions. In the absence of a chemical potential, if the imaginary parts of the complexified fields are initially
zero, $\phi^\rmI = 0$, the forces in the imaginary direction will be zero always, and therefore the configuration will 
remain constrained in the manner of real Langevin dynamics. This is called a \emph{cold start}. 
Alternatively, the imaginary parts of the fields may be 
initialised randomly according to a Gaussian distribution, called a \emph{hot start}. 

\begin{figure}[htpb]
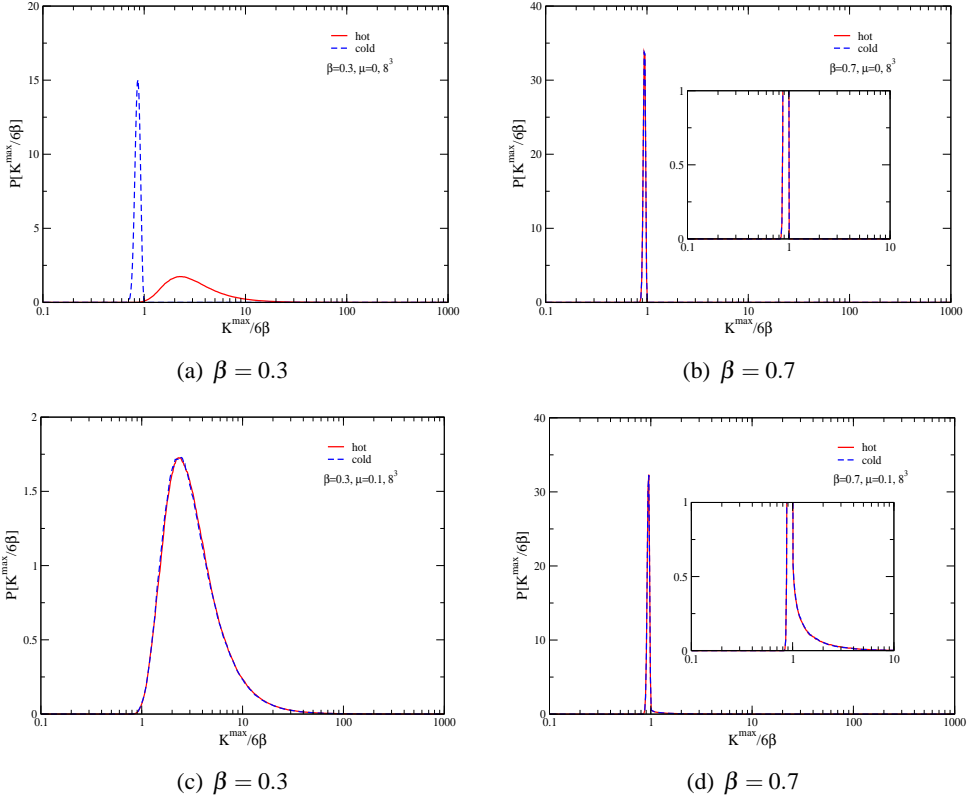

\centering
\subfigure[$\beta=0.3$]{
\includegraphics[totalheight=0.2\textheight, width=0.4\textwidth]{prob_kmax_b0.3_mu0.0_x8.eps}
}
\hspace{0.1in}
\subfigure[$\beta=0.7$]{
\includegraphics[totalheight=0.2\textheight, width=0.4\textwidth]{prob_kmax_b0.7_mu0.0_x8.eps}
\label{fig:kmax-plot-b0.7-mu0.0}
}
\vspace{0.1in}
\subfigure[$\beta=0.3$]{
\includegraphics[totalheight=0.2\textheight, width=0.4\textwidth]{prob_kmax_b0.3_mu0.1_x8.eps}
\label{fig:kmax-plot-b0.3-mu0.1}
}
\hspace{0.1in}
\subfigure[$\beta=0.7$]{
\includegraphics[totalheight=0.2\textheight, width=0.4\textwidth]{prob_kmax_b0.7_mu0.1_x8.eps}
\label{fig:kmax-plot-b0.7-mu0.1}
}
\caption{Distributions of $K^\rmMax$ using hot and cold starts for $\mu=0$ (top) and $\mu=0.1$ (bottom), lattice volume $8^3$, at two
$\beta$ values. Note the different vertical scales.}
\label{fig:kmax-plots}
\end{figure}

A good quantity for making such comparisons is the maximal force $K^\rmMax$. When $\phi^\rmI=0$, it will be constrained such that
$K^\rmMax \leq 6\beta$. With complexified dynamics, there is no upper bound and the drift terms can fluctuate over several orders of 
magnitude~\cite{Aarts:2009dg}. 
%
In Fig.~\ref{fig:kmax-plots} 
distributions of the maximal force term for high and low $\beta$ using hot and cold starts are plotted. 

In the absence of a chemical potential, the difference in behaviour between hot and cold starts is clear. At large $\beta$ the 
complex Langevin (hot start) configurations are driven to the same equilibrium distribution as the real Langevin (cold start) ones. 
The drift terms are bounded by $6\beta$ in both cases. However, at small $\beta$, 
the drift terms from the hot start configurations fluctuate over several orders of magnitude, whereas the drift terms from
the cold start are bounded by $6\beta$. Since the corresponding result for the action density is incorrect, this 
indicates that complex Langevin dynamics is not reaching a correct equilibrium distribution.
At nonzero chemical potential, $\mu=0.1$, the distributions coincide because, even in the case of a cold start, the chemical potential forces 
the field into the complexified space. Consequently, this problem is not caused by the initial conditions, but by the dynamics itself \cite{Aarts:2010aq}. 
At large $\beta$, there are occasional occurrences of large drift terms (Fig.~\ref{fig:kmax-plot-b0.7-mu0.1}, inset), but the distribution
is still dominated by the peak at $6\beta$. At small $\beta$, this peak is absent and the maximal drift terms are substantially larger 
(Fig.~\ref{fig:kmax-plot-b0.3-mu0.1}).

\section{Summary}
Complex Langevin dynamics offers the potential for a general solution to the sign problem. The XY model at finite chemical
potential was studied using complex Langevin dynamics and compared to results obtained with the alternative 
world line formulation, which is free of the sign problem. It is necessary to integrate the stochastic equations using a dynamic stepsize, 
in order to eliminate the problem of runaway solutions that appear when using a fixed stepsize. 
The action density $\langle S\rangle/\Omega$ was computed in the region around $\mu^2\sim0$. A comparison between results from 
using an imaginary chemical potential, $\mu^2<0$, with those at $\mu^2>0$ shows that complex Langevin dynamics gives the correct
result at large $\beta$ but an incorrect result at small $\beta$. An analysis of the relative difference between the action densities obtained using 
complex Langevin dynamics and the world line formalism over a wide region in the $\beta$-$\mu$ plane shows that the failure
of complex Langevin dynamics is strongly correlated with the phase of the theory. In the ordered phase at $\beta>\beta_{c}$
complex Langevin dynamics works correctly, but it fails in the disordered phase $\beta<\beta_{c}$.

To investigate this difference further, we computed the distribution of the maximal force term $K^\rmMax$ at $\mu=0$ using
two different initial conditions in which the imaginary parts of the fields were set randomly (hot) or to zero (cold). When complex Langevin dynamics
yields the correct result (at large $\beta$), the configuration from the hot start is driven to the same distribution as from the cold start. However,
where it fails (at small $\beta$), the distributions do not match. We conclude, therefore, that the failure of complex Langevin dynamics
is not caused by the sign problem, but rather by an incorrect exploration of configuration space. This conclusion is supported by results
that indicate that complex Langevin dynamics can evade the sign problem, see e.g. Refs.~\cite{Aarts:2008rr, Aarts:2008wh, Aarts:2010gr}.
An investigation into the incorrect exploration of complexified field space is currently in progress, building on the ideas
put forward in Ref.~\cite{Aarts:2009uq}.

\acknowledgments
Discussion and collaboration with Ion-Olimpiu Stamatescu and Erhard Seiler is greatly appreciated. This work has been supported in part by the EU 
Integrated Infrastructure Initiative HadronPhysics2 and STFC.

\end{document}